\documentclass{eas}

\usepackage{graphicx}

\newcommand{\be}{\begin{eqnarray}}
\newcommand{\ee}{\end{eqnarray}}

\begin{document}

\title{Review of Observational Evidence for Dark Matter in the Universe
and in upcoming searches for Dark Stars}
\runningtitle{Freese: Observational Evidence for Dark Matter \dots}

\author{Katherine Freese}
\address{Michigan Center for Theoretical Physics, University of Michigan,
Ann Arbor, MI 48109, USA}

\date{\today}

\begin{abstract}
Over the past decade, a consensus picture has emerged
in which roughly a quarter of the universe consists of
dark matter. The observational evidence for the existence of
dark matter is reviewed: rotation curves of galaxies, weak
lensing measurements, hot gas in clusters,
primordial nucleosynthesis and microwave background experiments.
In addition, a new line of research on Dark Stars is presented,
which suggests that the first stars to exist in the universe
were powered by dark matter heating rather than by fusion:
the observational possibilities of discovering dark matter in this
way are discussed.
\end{abstract}

\maketitle

\section{Introduction}
A standard model of cosmology is emerging (often dubbed the
Concordance Model), in which the universe consists of 4\% ordinary
baryonic matter, $\sim 23$\% dark matter, and $\sim 73$\% dark energy,
with a tiny abundance of relic neutrinos.  The baryonic content is
well-known, both from element abundances produced in primordial
nucleosynthesis roughly 100 seconds after the Big Bang, and from
measurements of anisotropies in the cosmic microwave background (CMB).
The evidence for the existence of dark matter is overwhelming, and
comes from a wide variety of astrophysical measurements.

\section{Dark Matter in Galaxies and Clusters}
The evidence that 95\% of the mass of galaxies and clusters is made of
some unknown component of Dark matter (DM) comes from (i) rotation
curves (out to tens of kpc), (ii) gravitational lensing (out to 200
kpc), and (iii) hot gas in clusters.

\subsection{Rotation Curves}

In the 1970s, Ford and Rubin  \cite{FordRubin1970} 
discovered that rotation curves of
galaxies are flat.  The velocities of objects (stars or gas) orbiting
the centers of galaxies, rather than decreasing as a function of the
distance from the galactic centers as had been expected, remain
constant out to very large radii.  Similar observations of flat
rotation curves have now been found for all galaxies studied,
including our Milky Way.  The simplest explanation is that galaxies
contain far more mass than can be explained by the bright stellar
objects residing in galactic disks.  This mass provides the force to
speed up the orbits.  To explain the data, galaxies must have enormous
dark halos made of unknown Òdark matter.Ó Indeed, more than 95\% of
the mass of galaxies consists of dark matter.  This is illustrated in
Fig. 1, where the velocity profile of galaxy NGC 6503 is displayed as
a function of radial distance from the galactic center. The baryonic
matter which accounts for the gas and disk cannot alone explain the
galactic rotation curve. However, adding a dark matter halo allows a
good fit to data.

The limitations of rotations curves are that one can only look out as
far as there is light or neutral hydrogen (21 cm), namely to distances
of tens of kpc.  Thus one can see the beginnings of DM haloes, but
cannot trace where most of the DM is. The lensing experiments
discussed in the next section go beyond these limitations.

\begin{figure}
\includegraphics[width=\textwidth]{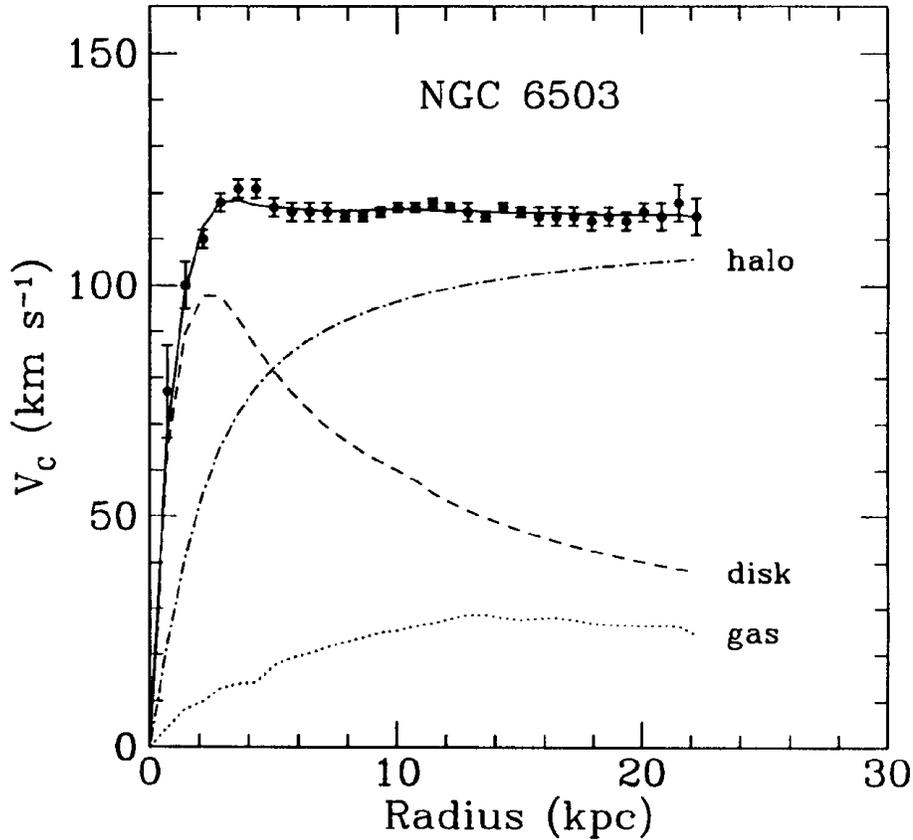}
\caption{Galactic rotation curve for NGC 6503 showing disk and gas
  contribution plus the dark matter halo contribution needed to match
  the data.}
\end{figure}

\subsection{Lensing}

Einstein's theory of General Relativity predicts that mass bends, or
lenses, light.  This effect can be used to gravitationally ascertain
the existence of mass even when it emits no light.  Lensing
measurements confirm the existence of enormous quantities of dark
matter both in galaxies and in clusters of galaxies.

Observations are made of distant bright objects such as galaxies or
quasars.  As the result of intervening matter, the light from these
distant objects is bent towards the regions of large mass.  Hence
there may be multiple images of the distant objects, or, if these
images cannot be individually resolved, the background object may
appear brighter.  Some of these images may be distorted or sheared.
The Sloan Digital Sky Survey used weak lensing (statistical studies of
lensed galaxies) to conclude that galaxies, including the Milky Way,
are even larger and more massive than previously thought, and require
even more dark matter out to great distances (Adelman-McCarthy {\em et al.\/}
\cite{Sloan2005}).  Again, the
predominance of dark matter in galaxies is observed.

A beautiful example of a strong lens is shown in Figure 2.  The panel
on the right shows a computer reconstruction of a foreground cluster
inferred by lensing observations made by Tyson et al. using the Hubble
Space Telescope.  This extremely rich cluster contains many galaxies,
indicated by the peaks in the figure.  In addition to these galaxies,
there is clearly a smooth component, which is the dark matter
contained in clusters in between the galaxies.

The key success of the lensing of DM to date is the evidence that DM
is seen out to much larger distances than could be probed by rotation
curves: the DM is seen in galaxies out to 200 kpc from the centers
of galaxies, in agreement with
N-body simulations.  On even larger Mpc scales, there is
evidence for DM in filaments (the cosmic web).

\begin{figure}
\includegraphics[width=\textwidth]{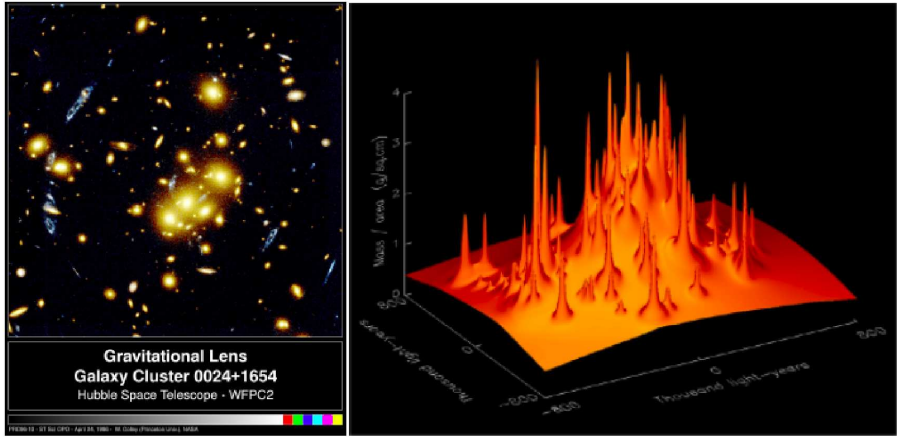}
\caption{Left: The foreground cluster of galaxies gravitationally
  lenses the blue background galaxy into multiple images. Right: A
 computer reconstruction of the lens shows a
  smooth background component not accounted for by the mass of the
  luminous objects.}
\end{figure}

\subsection{Hot Gas in Clusters}

Another piece of gravitational evidence for dark matter is the hot gas
in clusters.  Figure 3 illustrates the Coma Cluster. The left panel is
in the optical, while the right panel is emission in the x-ray
(observed by ROSAT)(Briel \& Henry \cite{Coma1997}).  
[Note that these two images are not
on the same scale.]  The X-ray image indicates the presence of hot
gas.  The existence of this gas in the cluster can only be explained
by a large dark matter component that provides the potential well to
hold on to the gas.

\begin{figure}
\includegraphics[width=0.5\textwidth]{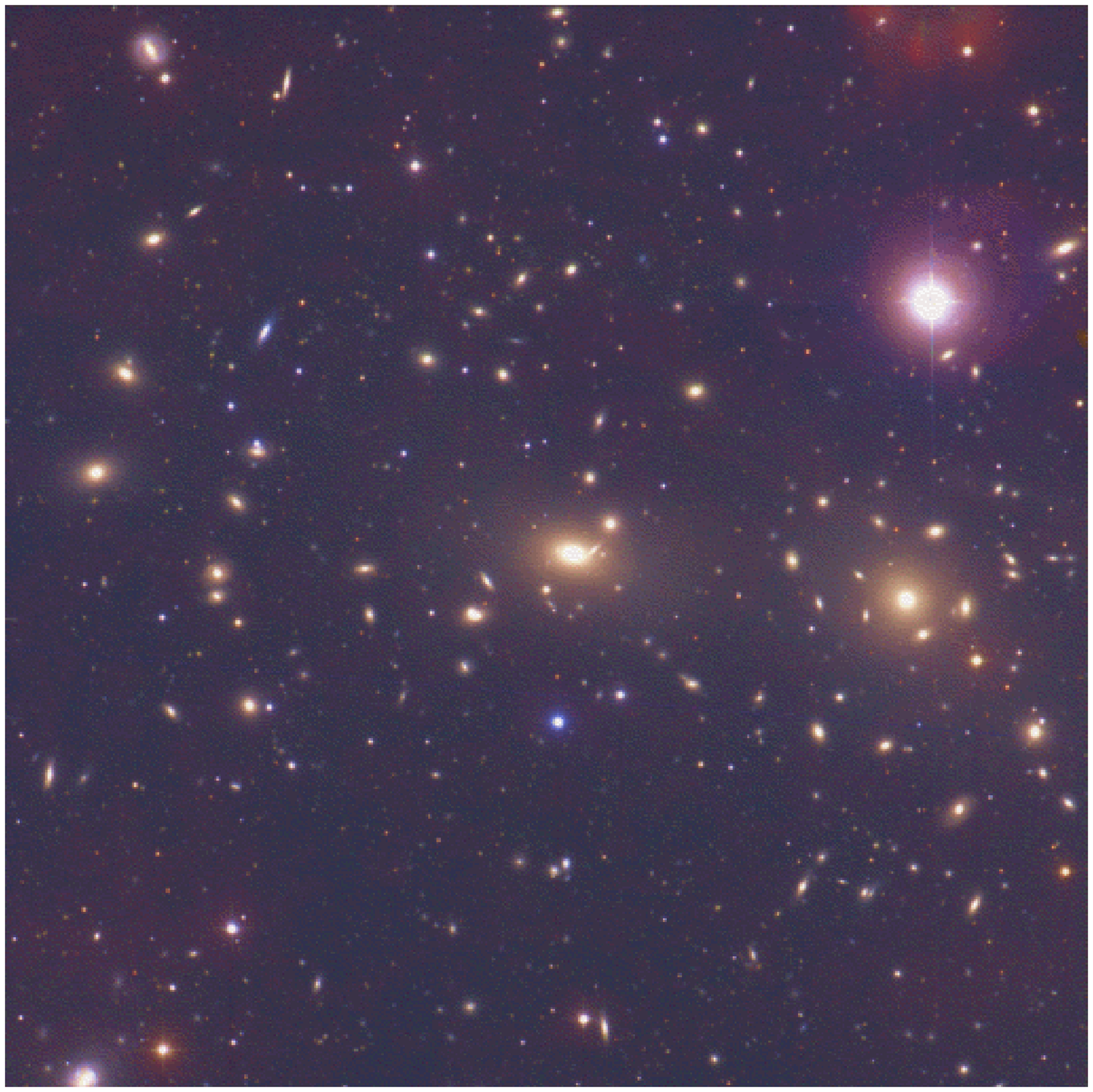}
\includegraphics[width=0.5\textwidth]{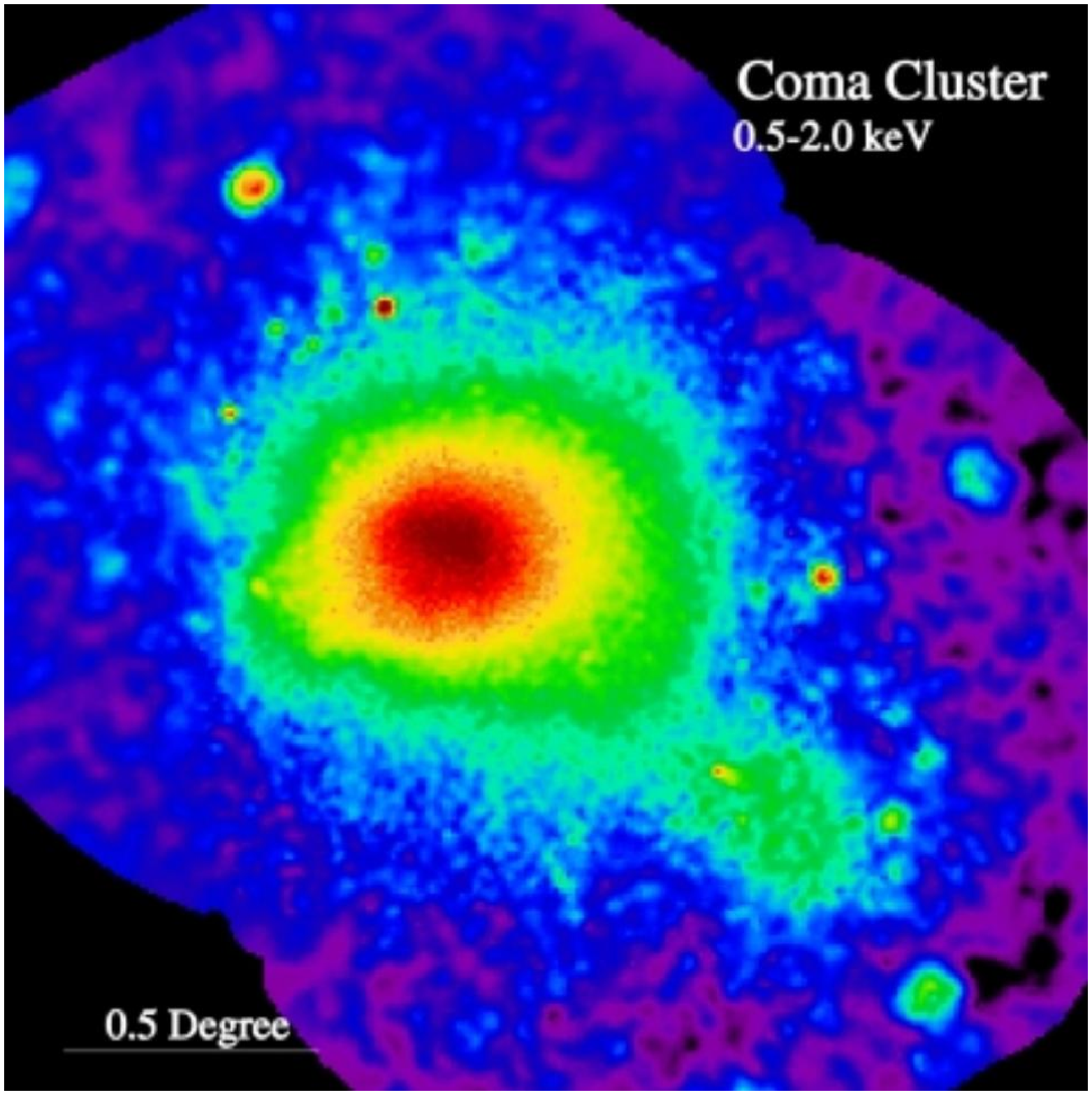}
\caption{COMA Cluster: without dark matter, the hot gas would
  evaporate. Left panel: optical image. Right panel: X-ray image from
  ROSAT satellite.}
\end{figure}

\subsection{Bullet Cluster}

A recent image of the bullet cluster of galaxies (a cluster formed out
of a collision of two smaller clusters) taken by the Chandra X-ray
observatory shows in pink the baryonic matter; in blue is an image of
the dark matter, deduced from gravitational lensing. In the process of
the merging of the two smaller clusters, the dark matter has passed
through the collision point, while the baryonic matter slowed due to
friction and coalesced to a single region at the center of the new
cluster. In modified gravity theories without dark matter, it is not
likely that such a differentiation of these two components of the
matter would take place.

\begin{figure}
\begin{center}
\includegraphics[width=7cm]{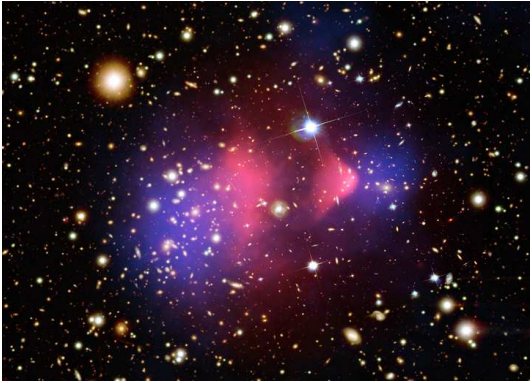}
\end{center}
\caption{A collision of galactic clusters (the bullet cluster) shows
  baryonic matter (pink) as separate from dark matter (blue), whose
  distribution is deduced from gravitational lensing.}
\end{figure}

In summary, the evidence is overwhelming for the existence of an
unknown component of DM that comprises 95\% of the mass in galaxies
and clusters.

\section{Cosmic Abundances}
The cosmic abundances tell a consistent story in which the
preponderance of the mass in the universe consists of an unknown DM
component.  The Cosmic Microwave Background provides the most powerful
measurements of the cosmological parameters; primordial
nucleosynthesis restricts the abundance of baryonic matter; Type IA
supernovae provided powerful evidence for the acceleration of the
universe, possibly explained by dark energy as the major constituent
of the cosmic energy density.

\subsection{The Cosmic Microwave Background}

Further evidence for dark matter comes from measurements on
cosmological scales of anisotropies in the CMB (WMAP Collaboration
\cite{wmap2003},\cite{wmap2008}).  The CMB is the remnant
radiation from the hot early days of the universe. The photons
underwent oscillations that froze in just before decoupling from the
baryonic matter at a redshift of 1000.  The angular scale and height
of the peaks (and troughs) of these oscillations are powerful
probes of cosmological parameters, including the total energy density,
the baryonic fraction, and the dark matter component.  The sound
horizon at last scattering provides a ruler stick for the geometry of
the universe: if the light travels in a straight line (as would be the
case for a flat geometry), then the angular scale of the first Doppler
peak was expected to be found at 1 degree; indeed this is found to be
correct.  Thus the geometry is flat, corresponding to an energy
density of the universe of $\sim 10^{-29} {\rm gm/cm}^3$.  The height
of the second peak implies that 4\% of the total is ordinary atoms,
while matching all the peaks implies that 23\% of the total is DM.

\subsection{Primordial nucleosynthesis}
When the universe was a few hundred seconds old, at a temperature of
ten billion degrees, deuterium became stable: $p + n \rightarrow D +
\gamma$.  Once deuterium forms, helium and lithium form as well. The
formation of heavier elements such as C, N, and O must wait a billion
years until stars form, with densities high enough for triple
interactions of three helium atoms into a single carbon atom. The
predictions from the Big Bang are 25\% Helium-4, $10^{-5}$
deuterium, and $10^{-10}$ Li-7 abundance by mass.  These predictions
exactly match the data as long as atoms are only 4\% of the total
constituents of the universe.

\subsection{Dark Energy}
Evidence for the 70\% dark energy in the universe comes from
observations of distant supernovae (Perlmutter {\em et al.\/}
\cite{sn1999a}, Riess {\em et al.\/} \cite{sn1999b}, Riess {\em et
  al.\/} \cite{sn2004}). The supernovae are dimmer than expected, as
is most easily explained by an accelerating universe.  There are two
different approaches to the dark energy: (i) a vacuum energy such as a
cosmological constant or time-dependent vacuum (Freese {\em et al.\/}
\cite{fafm1987}) may be responsible, or (ii) it is possible that
General Relativity is incomplete and that Einstein's equations need to
be modified (Freese \& Lewis \cite{modgenrel2002a}, Freese 2005
\cite{modgenrel2005}, Deffayet {\em et al.\/} \cite{modgenrel2002b},
Carroll {\em et al.\/} \cite{Carroll_etal2004}).  Note, however, that
this dark energy does not resolve or contribute to the question of
dark matter in galaxies, which remains as puzzling (if not more) than
twenty years ago.  We now have a concordance model of the universe, in
which roughly a quarter of its content consists of dark matter.

\section{Dark Matter Candidates}
 
 There is a plethora of dark matter candidates. 
MACHOs, or Massive Compact Halo Objects, are made of
ordinary matter in the form of faint stars or stellar remnants;
they could also be primordial black holes or mirror matter (Mohaptra
\& Teplitz \cite{MohapatraTeplitz1999}).
However, there are not enough of these to completely resolve the
question.  Of the nonbaryonic candidates, the most popular are the
WIMPS (Weakly Interacting Massive Particles) and the axions, as these
particles have been proposed for other reasons in particle physics.
Ordinary massive neutrinos are too light to be cosmologically
significant, though sterile neutrinos remain a possibility.  Other
candidates include primordial black holes, nonthermal WIMPzillas, and
Kaluza-Klein particles which arise in higher dimensional theories.

\subsection{MACHOs}

MACHO candidates include faint stars, planetary objects (brown
dwarfs), and stellar remnants (white dwarfs, neutron stars, and black
holes).  Microlensing experiments (the MACHO (Alcock {\em et al.\/}
\cite{alcock2000}) and EROS (Ansari {\em et al.\/}
\cite{eros2004}) experiments) as well as a combination of other
observational (HST) and theoretical results (Graff \& Freese 
\cite{graff1996}) have shown that MACHOs less massive than 0.1 $M_\odot$
make an insignificant contribution to the energy density of the Galaxy.  However, there is a detection
(Alcock {\em et al.\/} \cite{alcock2000}) of a roughly 20\% halo
fraction made of $\sim 0.5 M_\odot$ objects which might be made of
stellar remnants such as white dwarfs.  We found a number of
constraints: the progenitors produce observable element abundances
(C,N,He), they require an enormous mass budget, the initial mass
function must be extremely sharply peaked, and, most important, the
progenitors produce observable infrared radiation. Our conclusion from
these constraints is that at most 20\% of the Galactic Halo can be
made of stellar remnants (Freese {\em et al.\/}
\cite{Freese_etal2000}, Fields {\em et al.\/} \cite{ffgwpb},
Graff {\em et al.\/} \cite{ffgwpc}).

 \subsection{Axions}
\label{sec:axions}

The good news is that cosmologists don't need to ``invent'' new
particles.  Two candidates already exist in particle physics for other
reasons: axions and WIMPs.  Axions with masses in the range
$10^{-(3-6)}$ eV arise in the Peccei-Quinn solution to the strong-CP
problem in the theory of strong interactions.  Axion bounds (Asztalos
{\em et al.\/} \cite{rosenberg}) from the ADMX cavity experiment
are approaching the remaining parameter range.

\subsection{WIMPs (Weakly Interacting Massive Particles)}
\label{sec:WIMPs}

WIMPs are also natural dark matter candidates from particle physics.
These particles, if present in thermal abundances in the early
universe, annihilate with one another so that a predictable number of
them remain today.  The relic density of these particles comes out to
be the right value:
\begin{equation}
\Omega_\chi h^2 = (3 \times 10^{-26} {\rm cm}^3/{\rm sec})
/ \langle \sigma v \rangle_{ann}
\end{equation}
where the annihilation cross section $\langle \sigma v \rangle_{ann} $
of weak interaction strength automatically gives the right answer.
This coincidence is known as "the WIMP miracle" and is the reason why
WIMPs are taken so seriously as DM candidates.  The best WIMP
candidate is motivated by Supersymmetry (SUSY): the lightest
neutralino in the Minimal Supersymmetric Standard Model.
Supersymmetry in particle theory is designed to keep particle masses
at the right value.  As a consequence, each particle we know has a
partner: the photino is the partner of the photon, the squark is the
quark's partner, and the selectron is the partner of the electron.
The lightest superysmmetric partner is a good dark matter candidate
(see the reviews by Jungman  {\em et al.\/}\cite{Jungman_etal1996}, Lewin
\& Smith \cite{jkgb}, Primack  {\em et al.\/}\cite{jkgc}, 
Bertone {\em et al.\/}\cite{Bertone_etal2004}).

There are several ways to search for dark WIMPs. SUSY particles may be
discovered at the LHC as missing energy in an event.  In that case one
knows that the particles live long enough to escape the detector, but
it will still be unclear whether they
are long-lived enough to be the dark matter.  Thus
complementary astrophysical experiments are needed. In direct
detection experiments, the WIMP scatters off of a nucleus in the
detector, and a number of experimental signatures of the interaction
can be detected (Goodman \& Witten \cite{gw}, 
Drukier {\em et al.\/} \cite{dfs}).  In indirect detection experiments,
neutrinos are detected from the Sun or Earth that arise as
annihilation products of captured WIMPs; the first papers suggesting
this idea were by Silk {\em et al.\/} \cite{SOS} in the Sun; and
by Freese \cite{Freese1986} as well as Krauss, Srednicki and
Wilczek \cite{Krauss_etal1986} in the Earth.  Another way to detect WIMPs
is to look for anomalous cosmic rays from the Galactic Halo: WIMPs in
the Halo can annihilate with one another to give rise to antiprotons,
positrons, or neutrinos (Ellis {\em et al.\/} \cite{ellis}).  In addition,
neutrinos, Gamma-rays, and radio waves may be detected as WIMP
annihilation products from the Galactic Center (Gondolo \& Silk
\cite{gonsilk}).  Many talks in this conference will discuss ongoing
and planned DM searches.

\section{Dark Stars}

The first stars to form in the universe, at redshifts $z \sim 10-50$,
may be powered by dark matter annihilation for a significant period of
time (Spolyar, Freese, and Gondolo \cite{SpolyarFreeseGondolo08}). 
We have dubbed these objects ``Dark Stars.''

 As discussed in the last section, WIMP dark matter annihilation in the
early universe provides the right abundance today to explain the dark
matter content of our universe. This same annihilation process will
take place at later epochs in the universe wherever the dark matter
density is sufficiently high to provide rapid annihilation.  The first
stars to form in the universe are a natural place to look for
significant amounts of dark matter annihilation, because they form at
the right place and the right time. They form at high redshifts, when
the universe was still substantially denser than it is today, and at
the high density centers of dark matter haloes. 
 
The first stars form inside dark matter (DM) haloes of $10^6 M_\odot$
(for reviews see e.g.  Ripamonti \& Abel \cite{RipamontiAbel05},
Barkana \& Loeb \cite{BarkanaLoeb01}, and Bromm \& Larson
\cite{BrommLarson03}; see also Yoshida et al.  \cite{Yoshida_etal06}.)
One star is thought to form inside one such DM halo. The first stars
play an important role in reionization, in seeding supermassive black
holes, and in beginning the process of production of heavy elements in
later generations of stars.  It was our idea to ask, what is the
effect of the DM on these first stars?  We studied the behavior of
WIMPs in the first stars.  As our canonical values, we take $m_\chi =
100$GeV for the WIMP mass and $\langle \sigma v \rangle_{ann} = 3
\times 10^{-26} {\rm cm^3/sec}$ for the annihilation cross section
(motivated above).  We find that the annihilation products of the
dark matter inside the star can be trapped and
deposit enough energy to heat the star and prevent it from further
collapse.  A new stellar phase results, a Dark Star, powered
by DM annihilation as long as there is DM fuel.

\subsection{Three Criteria for Dark Matter Heating}

 WIMP annihilation produces energy at a rate per
unit volume 
\begin{equation}
  Q_{\rm ann} = \langle \sigma v \rangle_{ann} \rho_\chi^2/m_\chi
  \linebreak \simeq  10^{-29} {{\rm erg} \over {\rm cm^3/s}} \,\,\, {\langle
    \sigma v \rangle \over (3 \times 10^{-26} {\rm cm^3/s})} \left({n \over {\rm
        cm^{-3}}}\right)^{1.6} \left({100 {\rm GeV}\over m_\chi}\right) 
\end{equation}
where $\rho_\chi$ is the DM energy density inside the star and $n$ is
the stellar hydrogen density.  Paper I (Spolyar, Freese, \& Gondolo
\cite{SpolyarFreeseGondolo08}) outlined the three key ingredients
for Dark Stars: 1) high dark matter densities, 2) the annihilation
products get stuck inside the star, and 3) DM heating wins over other
cooling or heating mechanisms.  These same ingredients are required
throughout the evolution of the dark stars, whether during the
protostellar phase or during the main sequence phase.

{\bf First criterion: High Dark Matter density inside the star.}  Dark
matter annihilation is a powerful energy source in these first stars
because the dark matter density is high. To find the DM density
profile, we started with an NFW (Navarro, Frenk \& White 
\cite{NavarroFrenkWhite96}) profile for both DM and gas in the $10^6
M_\odot$ halo.  Originally we used adiabatic contraction ($M(r)r$ =
constant) (Blumenthal et al. \cite{Blumenthal_etal85}) and
matched onto the baryon density profiles given by Abel, Bryan \&
Norman \cite{AbelBryanNorman02} and Gao et
al. \cite{Gao_etal07} to obtain DM profiles; see also Natarajan,
Tan, \& O'Shea \cite{NatarajanTanO'Shea08} for a recent
discussion.  Subsequent to our original work, we have done an exact
calculation (which includes radial orbits) (Freese, Gondolo, Sellwood
\& Spolyar \cite{FreeseGondoloSellwoodSpolyar08}) and found that
our original results were remarkably accurate, to within a factor of
two.  At later stages, we also consider possible further enhancements
due to capture of DM into the star (discussed below).

{\bf Second Criterion: Dark Matter Annihilation Products get stuck
  inside the star}.  In the early stages of Pop III star formation,
when the gas density is low, most of the annihilation energy is
radiated away (Ripamonti Mapelli \& Ferrara
\cite{RipamontiMapelliFerrara06}). However, as the gas collapses
and its density increases, a substantial fraction $f_Q$ of the
annihilation energy is deposited into the gas, heating it up at a rate
$f_Q Q_{\rm ann}$ per unit volume.  While neutrinos escape from the
cloud without depositing an appreciable amount of energy, electrons
and photons can transmit energy to the core.  We have computed
estimates of this fraction $f_Q$ as the core becomes more dense. Once
$n\sim 10^{11} {\rm cm}^{-3}$ (for 100 GeV WIMPs), e$^-$ and photons
are trapped and we can take $f_Q \sim 2/3$.

{\bf Third Criterion: DM Heating is the dominant heating/cooling
  mechanism in the star}.  We find that, for WIMP mass $m_\chi =
100$GeV (1 GeV), a crucial transition takes place when the gas density
reaches $n> 10^{13} {\rm cm}^{-3}$ ($n>10^9 {\rm cm}^{-3}$).  Above
this density, DM heating dominates over all relevant cooling
mechanisms, the most important being H$_2$ cooling (Hollenbach
\& McKee \cite{HollenbachMcKee79}).

Figure 5 shows evolutionary tracks of the protostar in the
temperature-density phase plane with DM heating included
(Yoshida et al. \cite{Yoshida_etal08}), for two DM particle
masses (10 GeV and 100 GeV).  Moving to the right on this plot is
equivalent to moving forward in time.  Once the black dots are
reached, DM heating dominates over cooling inside the star, and the
Dark Star phase begins.  The protostellar core is prevented from
cooling and collapsing further.  The size of the core at this point is
$\sim 17$ A.U. and its mass is $\sim 0.6 M_\odot$ for 100 GeV mass
WIMPs.  A new type of object is created, a Dark Star supported by DM
annihilation rather than fusion.

\begin{figure}
\includegraphics[width=\textwidth]{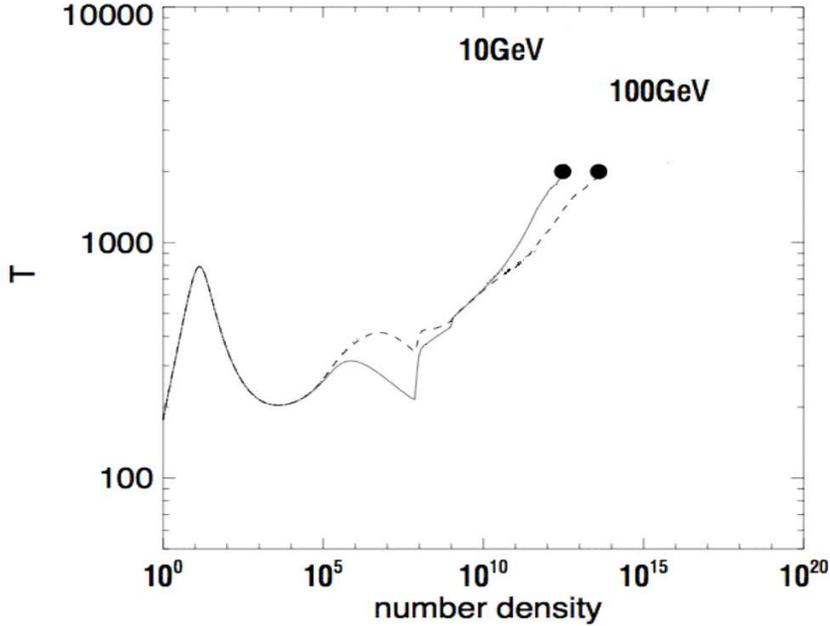}
\caption{ Temperature (in degrees K) as a function of hydrogen density
  (in cm$^{-3}$) for the first protostars, with DM annihilation
  included, for two different DM particle masses (10 GeV and 100 GeV).
  Moving to the right in the figure corresponds to moving forward in
  time.  Once the ``dots'' are reached, DM annihilation wins over H2
  cooling, and a Dark Star is created.}
\end{figure}

\subsection{Building up the Mass}

We have found the stellar structure of the dark stars
(hereafter DS) (Freese, Bodenheimer, Spolyar, \& Gondolo
\cite{FreeseBodenheimerSpolyarGondolo08}).  They accrete mass from the
surrounding medium.  In our paper we build up the DS mass as it grows
from $\sim 1 M_\odot$ to $\sim 1000 M_\odot$.
As the mass increases, the DS radius adjusts
until the DM heating matches its radiated luminosity.  We find
polytropic solutions for dark stars in hydrostatic and thermal
equilibrium. We build up the DS by accreting $1 M_\odot$ at a time
with an accretion rate of $2 \times 10^{-3} M_\odot$/yr, always
finding equilibrium solutions.  We find that initially the DS are in
convective equilibrium; from $(100-400) M_\odot$ there is a transition
to radiative; and heavier DS are radiative.  As the DS grows, it pulls
in more DM, which then annihilates.  We continue this process until
the DM fuel runs out at $M_{DS} \sim 800 M_\odot$ (for 100 GeV WIMPs).
Figure 6 shows the stellar structure. One can see ``the power of
darkness:'' although the DM constitutes a tiny fraction ($<10^{-3}$)
of the mass of the DS, it can power the star. The reason is that WIMP
annihilation is a very efficient power source: 2/3 of the initial
energy of the WIMPs is converted into useful energy for the star,
whereas only 1\% of baryonic rest mass energy is useful to a star via
fusion.

\begin{figure}
\includegraphics[width=0.5\textwidth]{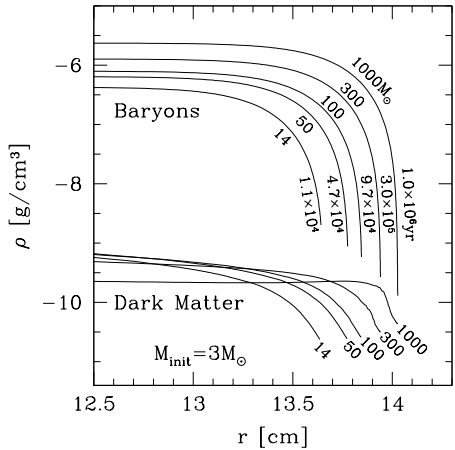}
\caption{Evolution of a dark star (n=1.5) as mass is accreted onto the
  initial protostellar core of 3 M$_\odot$.  The set of upper (lower)
  curves correspond to the baryonic (DM) density profile at different
  masses and times. Note that DM constitutes $<10^{-3}$ of the mass of
  the DS.}
\end{figure}

\subsection{Results and Predictions}

Our final result (Freese, Bodenheimer, Spolyar, \& Gondolo
\cite{FreeseBodenheimerSpolyarGondolo08}), is very large first
  stars; e.g., for 100 GeV WIMPs, the first stars have $M_{DS} = 800
M_\odot$.  Once the DM fuel runs out inside the DS, the star contracts
until it reaches $10^8$K and fusion sets in.  A possible end result of
stellar evolution will be large black holes.  The Pair
Instability SN (Heger \& Woosley \cite{HegerWoosley02}) that
would be produced from 140-260 $M_\odot$ stars (and whose chemical
imprint is not seen) would not be as abundant.  Indeed this process
may help to explain the supermassive black holes that have been found
at high redshift ($10^9 M_\odot$ BH at z=6) and are, as yet,
unexplained (Li et al. \cite{Li_etal07}; Pelupessy et
al. \cite{Pelupessy_etal07}).

 The stars are very bright,
$\sim 10^6 L_\odot$, and relatively cool, (6000-10,000)K (as opposed
to standard Pop III stars whose surface temperatures exceed $30,000K$).  
Reionization
during this period is likely to be slowed down, as these stars can
heat the surroundings but not ionize them.  
One can thus
hope to find DS and differentiate them from standard Pop III stars.

\subsection{Later stages: Capture}

The dark stars will last as long as the DM fuel inside them persists.  The
original DM inside the stars runs out in about a million years.
However, as discussed in the next paragraph, the DM may be replenished
by capture, so that the DS can live indefinitely due to DS
annihilation.  We suspect that the DS will eventually leave their high
density homes in the centers of DM haloes, especially once mergers of
haloes with other objects takes place, and then the DM fuel will run
out. The star will eventually be powered by fusion. Whenever it again
encounters a high DM density region, the DS can capture more DM and be
born again.

The new source of DM in the first stars is capture of DM particles
from the ambient medium.  Any DM particle that passes through the
DS has some probability of interacting with a nucleus in the star
and being captured. The new particle physics ingredient required
here is a significant scattering cross section between the WIMPs
and nuclei. Whereas the annihilation cross section is
fixed by the relic density, the scattering cross section is a
somewhat free parameter, set only by bounds from direct detection 
experiments.  
Two simultaneous papers (Freese, Spolyar, \& Aguirre
\cite{FreeseSpolyarAguirre08}, Iocco \cite{Iocco08}) found the
same basic idea: the DM luminosity from captured WIMPs can be larger
than fusion for the DS. Two uncertainties exist here: the scattering
cross section, and the amount of DM in the ambient medium to capture
from.  DS studies following the original papers that include
capture have assumed (i)  the maximal scattering cross sections allowed by
experimental bounds and (ii) ambient DM densities that are never depleted.
With these assumptions, DS evolution models with DM heating after the
onset of fusion have now been studied in several papers (Iocco
et al.  \cite{Iocco_etal08}, Taoso et al.  \cite{Taoso_etal08},
Yoon et al  \cite{Yoon_etal08}).

In short, the first stars to form in the universe may be Dark Stars
powered by DM heating rather than by fusion.  Our work indicates that
they may be very large ($800 M_\odot$ for 100 GeV mass WIMPs). Once DS
are found, one can use them as a tool to study the properties of WIMPs.

\section{Conclusion}
95\% of the mass in galaxies and clusters of galaxies is in the form
of an unknown type of dark matter. We know this from rotation curves,
from gravitational lensing, and from hot gas in clusters.  This
assessment of the DM contribution to the global energy density of the
universe is consistent with measurements of the cosmic microwave
background, primordial nucleosynthesis, supernova, and large scale
structure.  A consensus picture has emerged, in which the DM
contributes 23\% of the overall energy density of the universe.  Its
nature is still unknown.  At most 1/5 of the DM in galaxies can be
white dwarfs (or other MACHO candidates), but most is likely to be an
exotic particle candidate.  DM searches for the best motivated
candidates, axions and WMPs are ongoing and promising over the next
few years.  One of the key properties of WIMP candidates is its
annihilation cross section, yielding the proper relic density
today. As a consequence of this annihilation, the first stars in the
universe may provide another avenue to test the DM hypothesis. These
stars may be powered by DM annihilation, and one can look for them in
upcoming telescopes.  The goal is to decipher this unknown dark matter
over the next decade.

\end{document}